\documentclass[12pt,onecolumn]{article}
\usepackage[utf8]{inputenc}
\usepackage[margin=0.5 in]{geometry}
\usepackage{graphicx}
\usepackage{biblatex}
\usepackage{amsthm}
\newtheorem*{defn}{Definition}
\usepackage{multirow}
\usepackage{titling}
\usepackage{authblk}
\usepackage{float}
\usepackage[caption = false]{subfig}
\title{Study of Stylized Facts in Stock Market Data}                    
\author{Vaibhav Sherkar\textsuperscript{1}{\footnote{\textsuperscript{1}Indian Statistical Institute, Kolkata \
,Email: vaibhav.sherkar2002@gmail.com}}\,
, Dr. Rituparna Sen \textsuperscript{2}{\footnote{\textsuperscript{2}
Indian Statistical Institute, Bengaluru\ 
,Email: rsen@isibang.ac.in}} }
\date{}

\providecommand{\keywords}[1]
{
  \small	
  \textbf{\textit{Keywords---}} #1
}

\begin{document}
\maketitle

\begin{abstract}
    A property of data which is common across a wide range of instruments, markets and time periods is known as stylized empirical fact in the financial statistics literature. This paper first presents a wide range of stylized facts studied in literature which include some univariate distributional properties, multivariate properties and time series related properties of the financial time series data. In the next part of the paper, price data from several stocks listed on 10 stock exchanges spread across different continents has been analysed and data analysis has been presented.
\end{abstract}
\section{Introduction}
A financial market is a platform where various financial instruments are traded (i.e. bought and sold).
Financial time series data primarily includes trading data of various financial instruments like Stocks, Commodities, Currencies, etc.
This report focuses on data analysis of trading data of Stocks.\par
As mentioned by R.Cont \cite{cont} a stylized empirical fact is
\emph{a property of data which is common across a wide range of instruments ,markets and time periods.}
 Several Stylized empirical facts in financial time series data have been identified and studied.

Uncertainty and Risk are always involved in stock trading.Hence investors are always interested in prediction of prices of stocks in future. Several attempts to model the stock prices have been made. Given, the huge amount of money and higher degree of uncertainty involved in stock trading, modelling the stock prices have always been a topic of interest for not only investors but also regulators of the market.  \par
It is expected that the proposed model for prices should be able to explain the stylized facts observed in the financial time series data. In other words, stylized facts can be used to ‘test’ the model.
Hence, identification of new stylized facts and verification of already identified stylized facts is a crucial task for accurate prediction of returns of stocks. This report mainly focuses on verification of several identified stylized facts in the stock market data of 10 different stock exchanges situated in different countries. Since, less work has been done for verification of stylized facts in data of emerging markets compared to as that of developed markets, many of the emerging markets have been considered among the 10 markets chosen.\par
Articles \cite{cont}, \cite{Taylor}, \cite{UCL_article} were referred to know about various stylized facts commonly known in financial time series data.The book 'Analysis of financial time series' by Ruey S. Tsay\cite{Time_series} was referred for studying necessary concepts and theory of time series analysis. Data analysis for verification of various Stylized facts in markets of Peru\cite{Peru}, Nigeria\cite{Nigeria}, Morocco\cite{Morocco}, India\cite{Indian_market} were studied.
The data analysis for verification of many of the stylized facts presented in this project has been done along the lines of analysis presented in article by R.Sen and M.Subramaniam \cite{Indian_market}. \par
This paper is structured as follows: Section 2 provides a description and source of data used for data analysis. Various stylized facts studied in literature along with statistical methods used to verify/contradict the stylized fact in the data have been presented in section 3. Section 4 provides an insight into data analysis. Section 5 concludes the paper presenting the conclusions. 

\section{Description of Data}
Daily closing price and Volume data of number of constituents of leading index of the 10 chosen markets for 8 to 10 years in 2010-2019 has been considered for data analysis. Relevant information regarding the data has been presented in table 1.
\begin{table}[b!]
\centering
\begin{tabular}{ |p{2.25cm}|p{3cm}|p{3cm}|p{1.5cm}|p{2.8cm}|  }
\hline
Country Name & Stock Exchange Name & Index considered & Number of stocks considered for data analysis & Time period for which data is considered\\
\hline
Brazil & Sao Paulo Stock Exchange & Ibovespa & 50 & January 2010-December 2019\\
Canada & Toronto Stock Exchange & S\&P/TSX 60 & 55 & January 2010-December 2019\\
Chile  & Santiago Stock Exchange & S\&P IPSA Index & 24 & January 2010-December 2019\\
China  & Shanghai Stock Exchange & SSE50 & 38 & July 2011-December 2019\\
Indonesia & Indonesia Stock Exchange & IDX80 & 52 & January 2010-December 2019\\
Mexico & Mexican Stock Exchange & S\&P/BMV IPC & 26 & June 2011-December 2019\\
Poland & Warsaw Stock Exchange & WIG30 & 23 & June 2011-December 2019\\
South Africa & Johannesburg Stock Exchange & JSE Top 40 & 31 & January 2010-December 2019\\
Thailand & The Stock Exchange of Thailand & SET50 & 35 & January 2010-December 2019\\
Turkey & BORSA Istanbul & BIST100 & 44 & January 2010-December 2019\\
\hline
\end{tabular}
\label{table:1}
\caption{Information Regarding data}
\end{table}
All the datasets used for data analysis have been downloaded from Yahoo Finance(https://finance.yahoo.com/)\par
Stationary time series is a time series whose statistical properties do not change over time.Stationarity is a desirable property of time series for statistical analysis. Prices of stocks are often not stationary as they exhibit a trend or seasonal component. To overcome this issue, we have considered log returns for statistical analysis instead of stock prices. Log returns are defined as follows.
\newline
\begin{equation}
    r_{t}=log\left(\frac{p_{t}}{p_{t-1}}\right)
    \label{eqn:Daily Returns}
\end{equation}
\begin{center}
 where $r_{t}$ denotes the log returns at time t and $p_{t}$ denotes the prices at time t.
\end{center}

 \begin{figure}
     \centering
     \includegraphics[width=15cm]{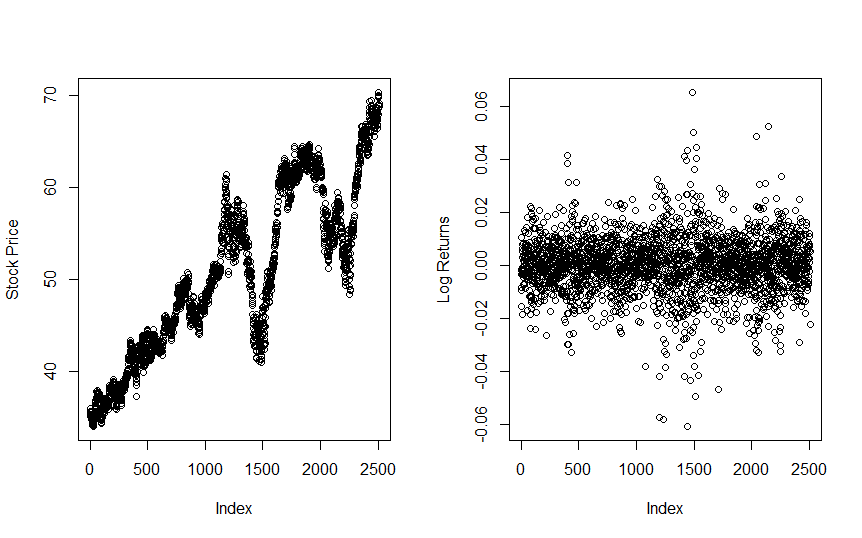}
     \caption{Stock Prices and Log Returns of 'TC Energy Corporation'}
     \label{fig:Stationarity}
 \end{figure}

Figure \ref{fig:Stationarity} represents the Price and log returns of the stock 'TC Energy Corporation' listed on Toronto Stock Exchange. By visual inspection, it is quite clear that concern of non-stationarity has been addressed properly by consideration of log returns instead of stock prices for statistical analysis.
\section{Stylized Empirical Facts}
\subsection{Univariate Distributional Stylized Empirical Facts}
\subsubsection{Gain Loss Assymetry}
\emph{one observes large drawdowns in stock prices and stock index values but not equally large upward movements.}\cite{cont}
\newline
According to this stylized fact, large number of observations in the time series of log returns are expected to be negative. 
Skewness of a random variable X is a measure of assymetry which is defined as follows
\begin{displaymath}
 \gamma_{1}=\frac{E(X-\mu)^3}{\sigma^3}
\end{displaymath}
\begin{center}
where $\mu$=E(X) and $\sigma^2$=Var(X)
\end{center}
Skewness is a measure of assymetry.
We calculate the sample skewness for the log returns of the stocks whose data is being considered. It is expected that skewness of log returns of the stocks is negative.

\subsubsection{Leverage Effect}
\emph{most measures of volatility of an asset are negatively correlated with the returns of that asset}\cite{cont}
\newline
One of the measures used to capture the volatility of a stock is Variance of returns over a certain time period.
Since, mean returns for many of the stocks are zero, we consider squared returns as a measure of volatility.
For random variables X and Y, we define correlation coefficient as
\begin{displaymath}
\rho=\frac{cov(X,Y)}{\sigma_{X}\sigma_{Y}}
\end{displaymath}
\begin{center}
where cov(X,Y)=E(XY)-E(X)E(Y),
 $\sigma_{X}^2$=Var(X), $\sigma_{Y}^2$=Var(Y)
\end{center}
We calculate the correlation coefficient between log returns and squared returns of same stock.
Positive correlation coefficient signifies positive association of two variables whereas negative correlation coefficient signifies the negative association of two variables.
According to stylized fact, the correlation coefficient between log returns and squared returns is expected to be negative.

\subsubsection{Aggregational Gaussinity}
\emph{as one increases the
time scale $\Delta$ t
over which returns are calculated,their distribution looks more and more like a normal distribution. In particular, the shape of the distribution
is not the same at different time scales.}\cite{cont}
\newline
Daily returns were calculated in eqn (\ref{eqn:Daily Returns}) with lag 1 in price of stock. Similarly,Weekly , Monthly and Quarterly returns were calculated with lag 5,20 and 60 in stock price respectively. 
According to the stylized empirical fact, the distribution should look more and more like normal distribution as we move from daily to quarterly returns.\par
\begin{figure}
     \centering
     \includegraphics[width=15cm]{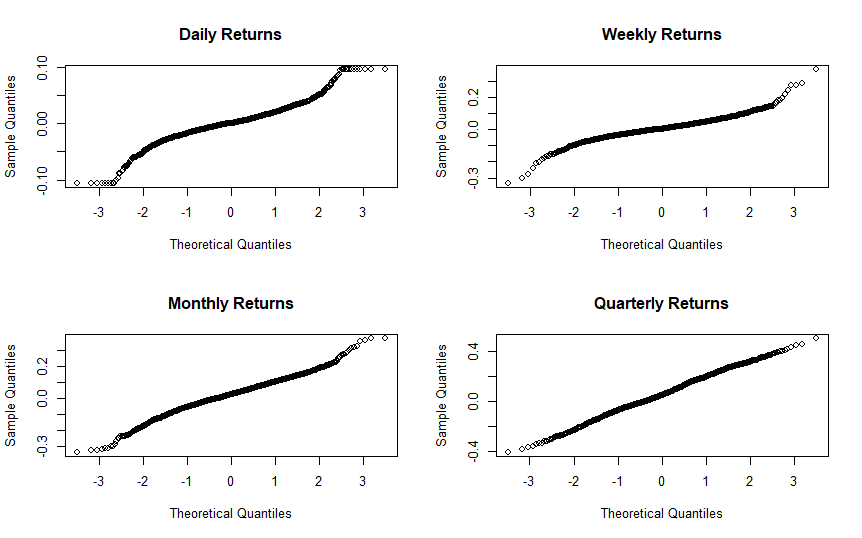}
     \caption{QQ plots of Daily, Weekly, Monthly and Quarterly returns against normal distribution for 'China Tourism Group Duty Free Corporation Limited' Stock}
     \label{fig:QQ plot}
 \end{figure}
By visual inspection of figure \ref{fig:QQ plot},which represents the QQ plots of daily, weekly, monthly and quarterly returns of 'China Tourism Group Duty Free Corporation Limited' stock which is listed on Shanghai Stock Exchange, it can be said that QQ plots looks more and more linear as we move from daily to quarterly returns which is interpreted as the distribution looks more and more like normal distribution as we move from daily to quarterly returns.\par
Kolmogorov-Smirnov test(KS),Shapiro-Wilke test(SW) and Jarque-Bera test(JB) were used to test the normality of daily,weekly,monthly and quarterly returns.p-values were recorded for each test conducted.According to the stylized fact, p-value is expected to increase as we move from daily to quarterly returns for same stock.

\subsubsection{Heavy Tails}
\emph{the (unconditional) distribution of return seems to display a power-law or Pareto-like tail, with a tail index which is finite, higher than two and less than five for most data sets studied}\cite{cont}
\newline
The pareto law is given by 
\begin{displaymath}
P[X>x]=\frac{k}{x^\beta}
\end{displaymath}
The pareto law is a special case of Power law.
Distribution of random varibales is usually studied in comparison with exponential distribution. Tail of a distribution is part of of distribution function when $|$X$|$ tends to infinity. The probability of occurence of extreme events is more in case of heavy tailed distributions as compared to the case of exponential distribution. Heavy tailed distributions are those distributions whose tail is not bounded by exponential distribution. Note that only part of distribution considered to determine heavy tailed or light tailed nature of distribution function is the tail of distribution function.\par
Let F(x) be Cumulative distribution function of a random variable X. Let $\bar{F}$(x)=1-F(x)
\newline
\begin{defn}
A function L: (0,$\infty$)$\rightarrow$ (0,$\infty$) is said to be \emph{slowly varying function} at infinity if 
\newline
\[ \lim_{x\to\infty} \frac{L(ax)}{L(x)} =1 ,\ \forall a>0\] 
\end{defn}
\begin{defn}
If $\exists$ L such  that
\newline
 \[\bar{F}(x)=x^{-{\frac{1}{\xi}}} L(x) \]
 \newline
where L is a slowly varying function at infinity, then $\xi$ is said to be tail
index of the distribution function F.
\end{defn}
Tail index for distributions of log returns were calculated using hill.adapt() function of Extremefit package of R. According to the stylized fact, the tail index of distribution of log returns is expected to lie in the interval [2,5].

\subsubsection{Decay of Distribution of volume as Power Law}
\emph{The distribution of Volume series decays as a power law.\cite{UCL_article}}
\newline
The tail index was calculated  using hill.adapt() function in extremefit package of R for volume series of stocks considered.According to the stylized fact, the tail index is expected to be finite.

\subsection{Multivariate Stylized Empirical Facts}
\subsubsection{Volume-Volatility Correlation}
\emph{trading volume is correlated with all measures of volatility}\cite{cont}
\newline
The correlation coefficient between log returns and trading volume has been calculated for every stock under consideration. According to the stylized fact,the correlation coefficient between log returns and trading volume is expected to be positive.

\subsubsection{Risk-Return Tradeoff}
\emph{Risk incurred in investment in a particular financial instrument and returns of that financial instrument are correlated }
\newline
Volatility of a stock has been considered as measure of risk. Here, the measure of volatility used is standard deviation of returns of a particular stock over full period of consideration. 
\newline
The correlation coefficient between mean return (Calculated over full period of consideration) and standard deviation of returns of stocks listed on a particular stock market has been calculated. According to the stylized fact, this correlation coefficient is expected to be positive for every market considered. 

\subsection{Time series Related Stylized Empirical facts}

\subsubsection{Asymmetry in Time Scales}
\emph{coarse-grained measures of volatility predict fine-scale volatility better than the other way round.}\cite{cont}
\newline
Article \cite{Asymmetry} was referred to understand the concepts presented in this subsection.
At first squared weekly returns were considered as a coarse grained measure of volatility and Variance of daily returns over a week was considered as fine scale volatility measure.
\begin{defn}
Let $X_{t}$ and $Y_{t}$ be two time series. Lagged correlation with lag h is defined as correlation coefficient between the time series $X_{t+h}$ and $Y_{t}$.
\end{defn}
Lagged correlations between fine-scale volatility measure and coarse-grained volatility measure were calculated for lag -10 to 10. Let $C_{l}$ denote the  lagged correlation coefficient with lag l. The difference $C_{l}-C_{-l}$ was calculated for l=1,2,...10 .According to stylized fact, this difference is expected to be positive. It was checked that if this difference is significantly positive for atleast one l. Same procedure was repeated by considering monthly squared return as a coarse grained measure of volatility and variance of daily returns over a month as fine scaled measure of volatility.
\begin{figure}
     \centering
     \includegraphics[width=15cm]{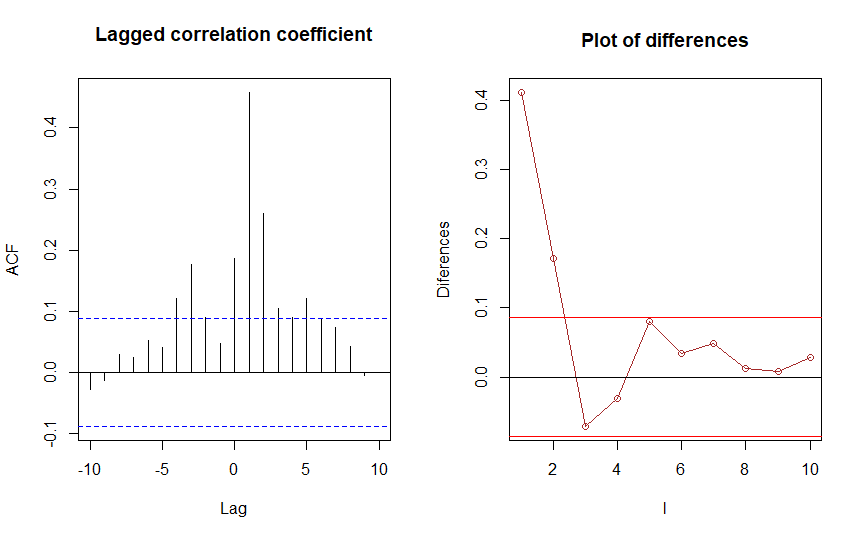}
     \caption{Lagged Autocorrelations and difference between autocorrelations at positive and negative lags for Thai Oil Public Company Limited Stock}
     \label{fig:CCF}
 \end{figure}

Left part of figure \ref{fig:CCF} shows the graph of lagged correlation coefficient between lags -10 to 10 for Thai Oil Public Company stock listed on the Stock Exchange of Thailand. The asymmetric nature of lagged corelation coefficients around lag 0 is quite clear from the this graph. The right part of figure \ref{fig:CCF} is plot of differences $C_{l}-C_{-l}$ on y-axis versus l on x-axis. Red lines show the level of significance with confidence coefficient 0.95. It is clearly visible that first two differences are significantly positive and rest are insignificant.

\subsubsection{Long Memory}
\emph{0.55 to 0.6 is the range of values of the ‘Hurst exponent’ reported in many studies of financial time series using the R/S or similar techniques} \cite{cont}
\newline
Article \cite{longmemory} was referred to understand the concept of long memory in financial time series, hurst exponent and R/S method to calculate hurst exponent.
\begin{defn}
Autocovariance function $\gamma$(k) for a time series $X_{t}$ is defined as
\[\gamma(k)=cov(X_{t+k},X_{t}) \]
\end{defn}
\begin{defn}
A time series is said to have long memory if its autocovariance function $\gamma$(k) satisfies
\[\gamma(k) \sim k^{-\alpha}L(k) \ as \ k \rightarrow \infty \]
where L is slowly varying function at infinity and 0$< \alpha <$1
\end{defn}
The exponent $\alpha$ is measure of long memory. Smaller is the $\alpha$, longer is the memory. 
\begin{defn}
Hurst exponent is defined as 
\[H=1-\frac{\alpha}{2}\]

\end{defn}
Hence long memory process exhibits H $>$ 0.5. Whereas short memory process ($\alpha=1$) has hurst exponent equal to 0.5 .
More and more H departs away from 0.5,higher is the long memory and hence higher is the predictability of the series. 
\newline
H greater than 0.5 indicates trend reinforcing series,showing a persistant behaviour. H less than 0.5 indicates the anti persistant behaviour i.e after a period of increase, the data tends to decrease and vice versa.
H always lies in the range 0 to 1 where white noise is characterized by 0 and linear trend is characterized by 1.
\newline
R/S method is used for estimation of hurst exponent. The original series of length T is divided into m subseries of length l.Mean and standard deviation for every subseries is computed.Now a different set of subseries is considered which is obtained by subtracting mean of subseries from every element of subseries for all subseries. In other words, we now consider subseries of deviations from mean.Now,a new subseries is computed corresponding to every subseries which is cumulative sum of deviation subseries. The range for every member of this set of subseries is calculated. R/S statistic is obtained for every  subseries by dividing the range of cumulative sum of deviations subseries by standard deviation of original subseries.Finally,mean of all  R/S statistics is obtained which is denoted as $(R/S)_{l}$
Now it is known that R/S statistic follows following relation asymptotically.

\[(R/S)_{l} \approx c l^{H}\]
$(R/S)_{l}$ statistic is calculated for different values of l.
Then hurst exponent is calculated by fitting a linear regression model between $(R/S)_{l}$ and log(l) using least squares method.
\[log((R/S)_{l}) = log(c) +Hlog(l)\]
The hurst exponent has been calculated using hurstexp() function in pracma package of R.
According to the stylized fact, the hurst exponent for return series is expected to lie between 0.55 to 0.6

\subsubsection{Long memory in volume series}
\emph{Volume series does exhibit long memory.\cite{UCL_article}}
\newline
Hurst exponent of volume series of stocks was calculated using hurstexp() function in pracma package of R. According to stylized fact, hurst exponent of volume series of stocks is expected to be gretar than 0.5 .

\subsubsection{Slow decay of autocorrelation in absolute returns}
\emph{the autocorrelation function of absolute returns decays slowly as a function of the time lag, roughly as a power law with an exponent $\beta$ which lies in the interval [0.2, 0.4]. This is sometimes interpreted as a sign of long-range dependence.} \cite{cont}
\newline
\begin{defn}
Autocorrelation of time series $X_{t}$ of order l is defined as
\[\rho_{l}=\frac{cov(X_{t},X_{t+l})}{Var(X_{t})}=\frac{\gamma(l)}{\gamma(0)}\]
\end{defn}
A power law model is fitted to autocorrelations of absolute returns of every stock and value of $\beta$ is found for every stock.
Let ac(l) be the autocorrelation function of absolute returns of a stock. The power law i.e $ac(l)=kl^{\alpha}$ was fitted to autocorrelation function of absolute returns.
\newline
\[log(ac(l))=log(k)+\alpha log(l)\]
\newline
Hence a linear regression model was fitted to the data of log of autocorrelation function of absolute returns of a stock and log of lag l using least squares method.
Exponent of power law has been defined as $-\alpha$. Exponent of power law ($\beta$) was calculated in each case.
Accoding to stylized fact $\beta$ is expected to lie in the interval [0.2,0.4].

\subsubsection{Absence of autocorrelations}
\emph{(linear) autocorrelations of asset returns are often insignificant, except for very small intraday time scales ($\approx$ 20 minutes) for which microstructure effects come into play.}\cite{cont}
\begin{figure}
     \centering
     \includegraphics[width=13cm]{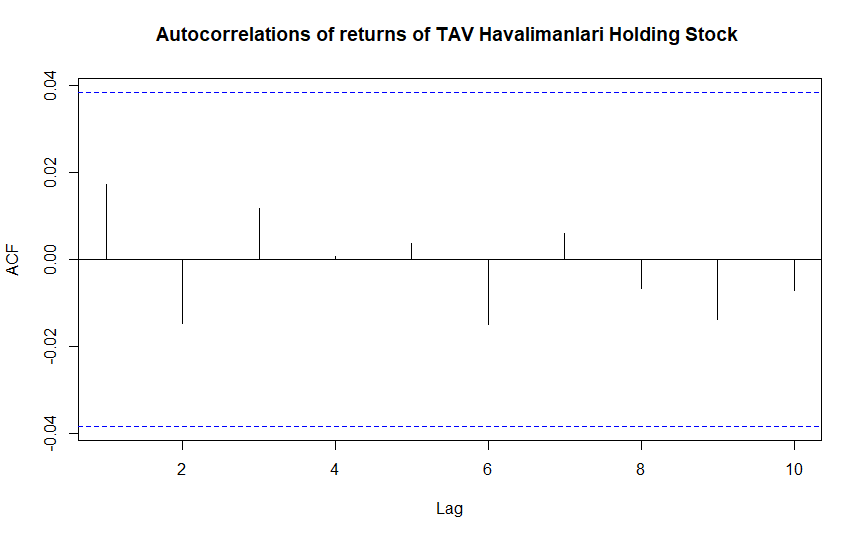}
     \caption{Autocorrelations of returns of 'TAV Havalimanlari Holding' Stock}
     \label{fig:ACF}
 \end{figure}
\newline
Figure \ref{fig:ACF} is plot of autocorrelations upto lag 10 of returns of stock 'TAV Havalimanlari Holding ' listed on BORSA Istanbul. Blue lines drawn in the graph are level of significance. It is clearly visible that all the autocorrelations upto lag 10 are insignificant for this stock as per the stylized fact. \par
For formal evaluation, portmanteau tests were conducted. Portmanteau tests are used to test the following:
\[\textbf{$H_{0}$:} \, r_{1}=r_{2}=....r_{m}=0 \] against
\[\textbf{$H_{1}$:} r_{i}\neq 0\ for \ atleast\ one \ i\ \in \{1,2,....m\} \]
Here $r_{i}$ is observed autocorrelation coefficient.
Two types of portanteau tests are used to test the above Hypothesis
\begin{description}
    \item[Box and pierce] 
    \[Q(m)=n \sum_{i=1}^{m} r_{i}^2\]
    
    \item [Ljung and Box]
    \[ Q(m)=n(n+2) \sum_{i=1}^{m} \frac{r_{i}^2}{n-i}\]
    Both the statistics follows asymptotic chi-square distribution with m degrees of freedom.
\end{description}
According to stylized fact, the null hypothesis in portmanteau tests is not expected to get rejected. 
\subsubsection{Volatility Clustering}
\emph{different measures of volatility display a positive autocorrelation over several days, which quantifies the fact that high-volatility events tend to cluster in time.}\cite{cont}
\newline
According to the stylized fact, period when large changes in returns occur are followed by large changes in returns and period when small changes occur are followed by small changes. We consider squared returns as a measure of volatility.
\begin{figure}
    \centering
    \includegraphics[width=15cm]{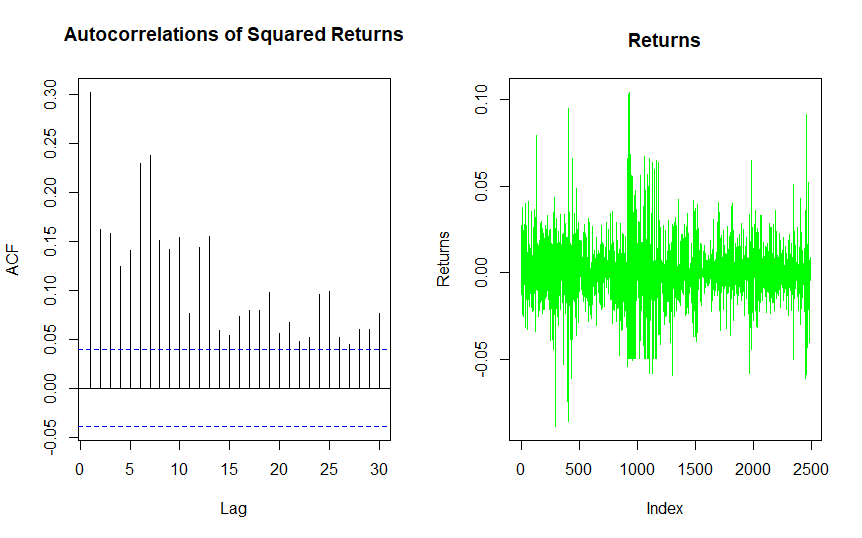}
    \caption{Autocorrelation of squared returns and returns of Banco de Credito e Inversiones}
    \label{fig:Vol_cluster}
\end{figure}
Left plot in figure \ref{fig:Vol_cluster} shows the autocorrelations of squared returns upto lag 30 of the 'Banco de Credito e Inversiones' stock listed on Santiago stock exchange. Blue dotted lines in the plot represents the level of significance. All the autocorrelations are significantly non zero as expected. Right plot in figure \ref{fig:Vol_cluster} shows the returns of same stock.It can be easily seen that the period of large changes in returns is followed by large changes in returns and period of small changes in returns is followed by small changes.\par
For formal calculation, portmanteau tests were carried out with null hypothesis that autocorrelations of squared returns upto lag 5 are zero. Two different tests to check if autocorrelation of squared returns of certain lag is 0 were carried out using the statistic $\frac{r\sqrt{n}}{1-r^2}$ and $\frac{r\sqrt{n-2}}{\sqrt{1-r^2}}$. Here, n represents number of observations and r represents observed autocorrelation coefficient of certain lag.The first statistic follows asymptotic normal distribution where the second statistic follows t distribution with $n-2$ degrees of freedom. According to stylized fact, null hypothesis in all the tests is expected to be rejected.   

\subsubsection{Conditional Heavy tails}
\emph{even after correcting returns for volatility clustering (e.g. via GARCH-type models), the residual time series still exhibit heavy tails. However, the tails are less heavy than in the unconditional distribution of returns.} \cite{cont}
\newline
Usually, time series of returns \{$r_{t}$\} is modelled as 
\begin{displaymath}
r_{t}=\mu_{t}+\epsilon_{t}
\end{displaymath}
Here $\mu_{t}$ denotes the conditional expectation i.e. E[$r_{t}|r_{t-1},r_{t-2},...$].
Autoregressive model is a very simple model used to model the time series. In this model $epsilon_{t}$ is assumed to be White Noise i.e. \{$\epsilon_{t}$\} are independent and identically distributed(iid) random variables with finite mean and variance.In AR(p) model, $\mu_{t}$ is represented as 
\begin{displaymath}
\mu_{t}=\phi_{0}+\phi_{1}r_{t-1}+....+\phi_{p}r_{t-p}
\end{displaymath}

Another common model used to model the time series is Autoregressive Moving Average(ARMA) model.
In ARMA model too \{$\epsilon_{t}$\} is assumed to be a White Noise innovation.In ARMA(p,q) model the conditional mean $\mu_{t}$ is represented as
\begin{displaymath}
\mu_{t}=\alpha_{0}+\alpha_{1}r_{t-1}+...\alpha_{p}r_{t-p} +\beta_{1}r_{t-1}+....+\beta_{q}r_{t-q}
\end{displaymath}
In these models,it can be seen that the conditional mean depends on t but conditional and unconditional variance is independent of t. To enable modelling time series data with non-constant volatility, Autoregressive Conditional Heteroscedastic(ARCH) model was propoposed.
Here $\epsilon_{t}$ is decomposed as $\epsilon_{t}=\sigma_{t}W_{t}$ and $W_{t}$ is assumed to be White noise and $\sigma_{t}$ is the standard deviation of $r_{t}$. In the model ARCH(p) $\sigma^2_{t}$ is modelled as
\begin{displaymath}
\sigma^2_{t}=\alpha_{0}+\alpha_{1}\epsilon^2_{t-1}+.....+\alpha_{p}\epsilon^2_{t-p}
\end{displaymath}
Here, all the coefficients are assumed to be non negative($\alpha_{0}$ is strictly positive.).
\newline
In reality,positive and negative shocks have different effects on Volatility. However,this fact cannot be explained by ARCH model. Hence, a modification to ARCH model was suggested. 
Another model used to model the volatility is Generalized ARCH(GARCH) model.
In GARCH(p,q) model, volatility $\sigma^2_{t}$ is modelled as followes
\begin{displaymath}
\sigma^2_{t}=\beta_{0}+\beta_{1}\epsilon^2_{t-1}+......\beta_{p}\epsilon^2_{t-p}+\gamma_{1}\sigma^2_{t-1}+.....+\gamma_{q}\sigma^2_{t-q}
\end{displaymath}
A GARCH(1,1) model is fitted to data of daily returns of stocks. Residuals were computed. According to stylized fact, residuals are expected to have heavy tails but less heavy than in unconditional distribution of returns.

\subsubsection{Intermittancy}
\emph{returns display, at any time scale, a high degree of variability. This is quantified by the presence of irregular bursts in time series of a wide variety of volatility estimators.}\cite{cont}
\newline
\begin{defn}
Kurtosis for a random variable X is defined as
\[K=\frac{E(X-\mu)^4}{\sigma^4}\]

\end{defn}

\begin{center}
    where $\mu=E(X)$ and $\sigma^2=Var(X)$
\end{center}
Intermittency can be characterized by high kurtosis.Kurtosis was computed for the return series and also for residual series obtained after fitting a GARCH model(to eliminate the time series effect from the data).
The value of kurtosis for normal distribution is 3. The one sided test to check if kurtosis is 3  against the alternative hypothesis that kurtosis is greter than was carreied out using the test statistic $\frac{\sqrt{n}(K-3)}{\sqrt{24}}$ which follows asymptotic normal distribution.
p-values were computed in each case.
Acoording to the stylized fact, the p-values are expected to be small i.e the null hypothesis that kurosis is 3 is expected to get rejected.

\subsubsection{Taylor effect}
\emph{For observed returns series the autocorrelation of absolute returns tends to be greater than the autocorrelation of squared returns.However, the term Taylor effect is also often used to refer to the more general result that the first order autocorrelation of $|R_{t}|^{d}$ is maximised when d=1, where $R_{t}$ is the observed return series}\cite{Taylor}
\newline
Observed autocorrelation of absolute returns and autocorrelation of squared returns were comapred for all stocks.
The value of d for which autocorrelation of $|R_{t}|^{d}$ is maximized was calculated using bisection and Newton Raphson method for every stock considered. Acoording to stylized fact, the numerical solution is expected to lie close to 1 for every stock considered.

\section{Results}
\subsubsection{Gain loss Assymetry}
\begin{figure}
    \centering
    \includegraphics[width=15cm]{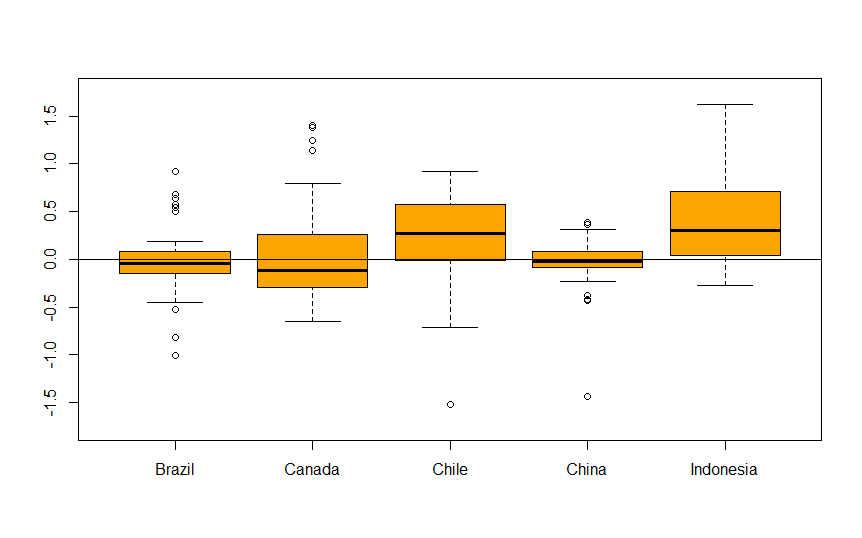}
    \caption{Observed values of skewness in markets of Brazil,Canada,Chile,China and Indonesia}
    \label{fig:skew_1}
\end{figure}
\begin{figure}
    \centering
    \includegraphics[width=15cm]{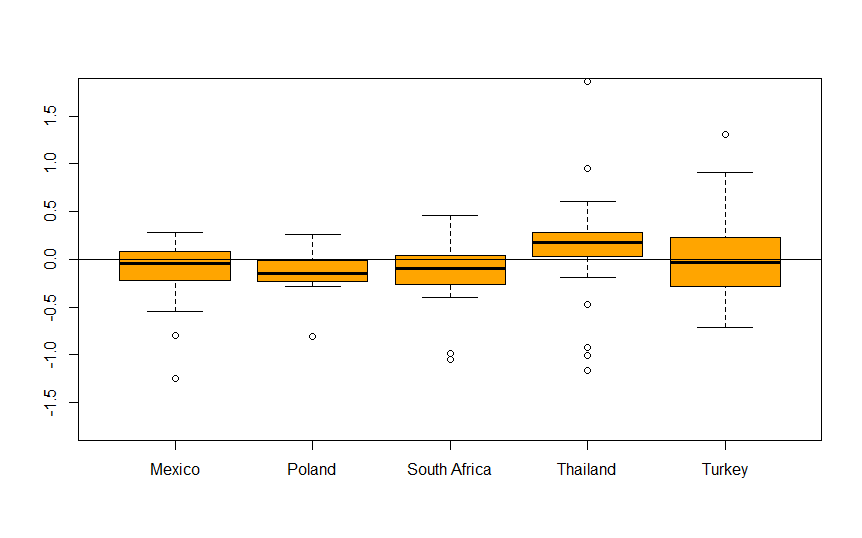}
    \caption{Observed values of skewness in markets of Mexico,Poland,South Africa,Thailand and Turkey}
    \label{fig:skew_2}
\end{figure}
Figure \ref{fig:skew_1} and \ref{fig:skew_2} represents the boxplots of skewness values for different markets in the range -1.75 to 1.75. In case of Chile,Indonesia and Thailand more than 75\% of the observed skewness values are positive similar to case of Indian market as mentioned in article \cite{Indian_market}contradicting the stylized fact. In case of Poland, 75\% of the observed skewness values are negative prociving evidence for validation of Stylized fact in Polish Stock Market. A large proportion of obserevd skewness values are negative in case of Canada(65.45\%) and South Africa(64.52\%) as expected. Observed skewness values are symmetric around zero in case of Brazil,China,Mexico and Turkey.
\subsubsection{Leverage Effect}
\begin{table}
\centering
\begin{tabular}{ |p{5cm}|p{7.5cm}|  }
\hline
Country Name & Proportion of stocks for which negative correlation between returns and squared returns is observed\\
\hline
Brazil & 0.5\\
Canada & 0.6\\
Chile & 0.25\\
China & 0.37\\
Indonesia & 0.12\\ 
Mexico & 0.54\\
Poland & 0.70\\
South Africa & 0.58\\
Thailand & 0.17 \\
Turkey & 0.5\\
\hline
\end{tabular}
\label{table:2}
\caption{Proportion of stocks which have negative correlation between returns and squared returns}
\end{table}
Table 2 shows the proportion of stocks which have negative correlation between returns and squared returns in each market. Since the proportion in case of Poland is much greater, it can be said that Poland exhibits strong leverage effect.The proportion is higher in case of Mexico,South Africa and Canada. However,Chile,China,Thailand and Indonesia show much lower proportion. Hence it can be said that leverage effect is reversed in markets of Chile,China,Thailand and Indonesia which is same as Indian market as mentioned in the article \cite{Indian_market}
\subsubsection{Aggregational Gaussinity}

\begin{figure}
     \centering
     \includegraphics[width=15cm]{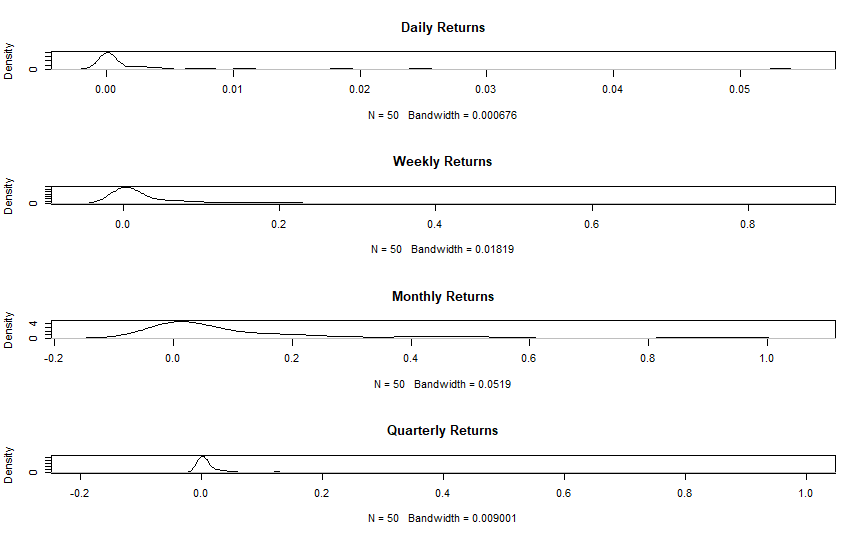}
     \caption{Densities fitted to p-values of KS test performed on returns of stocks in Market of Brazil}
     \label{fig:Gauss_Brazil}
 \end{figure}

Figure \ref{fig:Gauss_Brazil} shows the Kernel density estimates of p-values of KS test performed on daily, weekly, monthly and quarterly returns of stocks in Brazilian stock market.
The figures for kernel density estimates for other markets considered are similar and available on request.According to the stylized fact, p-values are expected to increase i.e. the peak of the estimated density is expected to shift rightwards. Peaks of densities do shift right as we move from daily to monthly returns in all the cases. However, the peak shifts left as we move from monthly to quarterly returns in all the markets except Thailand, Chile, Indonesia and Turkey. In case of turkey, the position of peak in estimated density for monthly and quarterly returns remains almost same. Hence, aggregational gaussinity has been verified as we move from daily to monthly returns but contradicted as we move from monthly to quarterly returns in most of the markets considered.

\subsubsection{Heavy Tails}
\begin{figure}
    \centering
    \includegraphics[width=15cm]{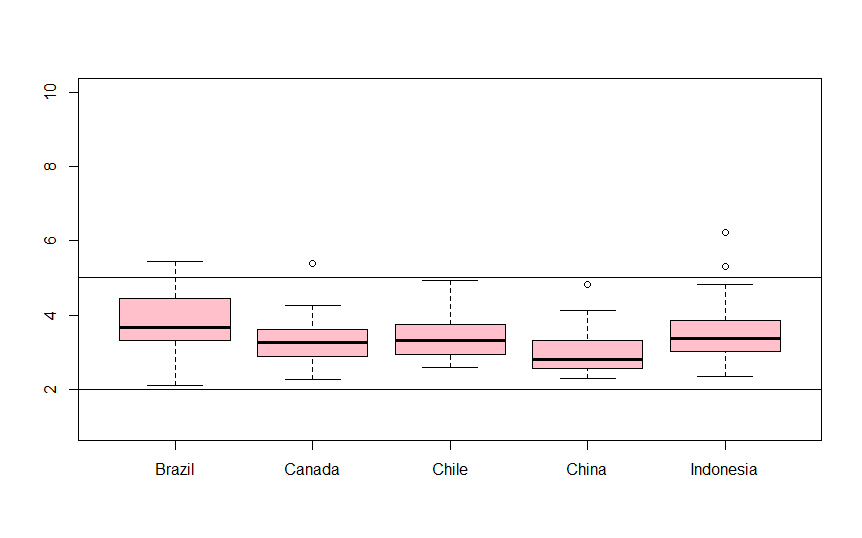}
    \caption{Observed values of Tail index in markets of Brazil,Canada,Chile,China and Indonesia}
    \label{fig:Tail_1}
\end{figure}
\begin{figure}
    \centering
    \includegraphics[width=15cm]{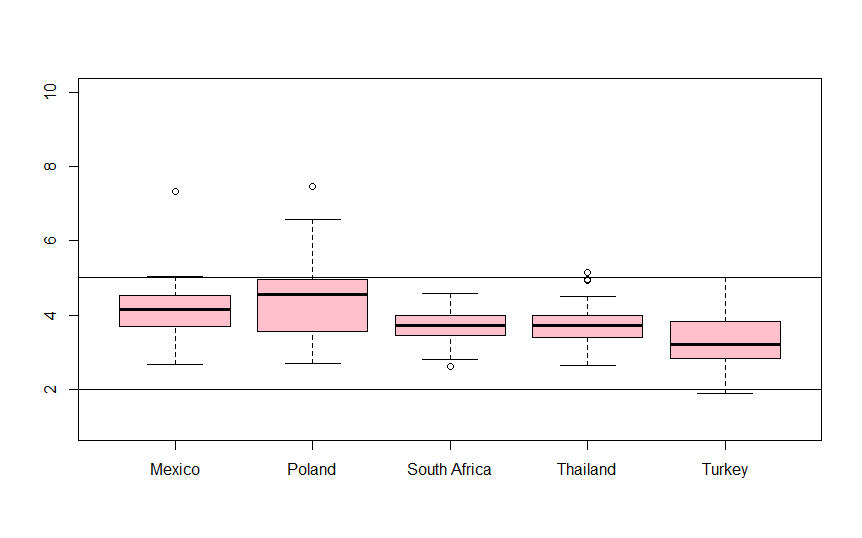}
    \caption{Observed values of Tail index in markets of Mexico,Poland,South Africa,Thailand and Turkey}
    \label{fig:Tail_2}
\end{figure}

In case of China, abnormally higher value of tail index was observed for 1 stock (Obsereved tail index was 38) and 5 stocks did not have heavy tails.
Figure \ref{fig:Tail_1} and \ref{fig:Tail_2} represents the tail index of returns of stocks in different markets in the range 1 to 10. Tail index less than 2 has been observed only in case of Turkey whereas Tail indices greater than 5 have been observed in all the markets except Chile,South Africa and Turkey.However, more than 75\% of the observed tail indices lie in the interval 2 to 5 in all the markets as expected.
\subsubsection{Decay of distribution of volume as power law}
Tail index for all the volumes series of the stocks considered is finite as expected.
\subsubsection{Volume-Volatility Correlation}
One negative value of correlation coefficient between volume and squared returns whose absolute value is much small (less than 0.02) has been obserevd in markets of Brazil,Chile and Indonesia.All the other values of observed correlation coefficients are positive as expected.
\subsubsection{Risk-Return Tradeoff}
\begin{table}
\centering
\begin{tabular}{ |p{5cm}|p{7.5cm}|  }
\hline
Country & correlation coefficient between mean and standard deviation of daily returns\\
\hline
Brazil & -0.59\\
Canada & -0.33\\
Chile & -0.70\\
China & 0.25\\
Indonesia & 0.19\\
Mexico & -0.42\\
Poland & 0.28\\
South Africa & -0.42\\
Thailand & 0.33\\
Turkey & 0.12\\
\hline
\end{tabular}
\label{table:3}
\caption{Correlation coefficient between standard deviation of daily returns and mean daily returns of all stocks in certain market}
\end{table}
Table 3 enlists the correlation coefficients between mean and standard deviation of daily returns of stocks listed on a certain stock exchange.Five countries exhibit negative correlation coefficient whereas other five exhibit positive coefficient.Note that all the asian countries exhibit positive correlation coefficient as expected.

\subsubsection{Asymmetry in Time Scales}
\begin{table}
\centering
\begin{tabular}{ |p{3cm}|p{7.5cm}|p{7.5 cm}|  }
\hline
Country & Proportion when weekly returns are considered & Proportion when monthly returns are considered\\
\hline
Brazil & 0.88 & 0.78\\
Canada & 0.85 & 0.84\\
Chile & 1 & 0.58\\
China & 1 & 0.89\\
Indonesia & 0.94 & 0.69\\
Mexico & 0.92 & 0.62\\
Poland & 0.91 & 0.83\\
South Africa & 0.90 & 0.90\\
Thailand & 0.94 & 0.91\\
Turkey & 0.91 & 0.86\\

\hline
\end{tabular}
\label{table:4}
\caption{Proportion of stocks having significant positive difference in $C_{l}$ and $C_{-l}$ }
\end{table}
By looking at the high values of proportion of stocks showing the asymmetry in time scales listed on the various stock exchanges considered, it can be said that the presence of stylized fact has been verified.

\subsubsection{Long Memory}
\begin{figure}
    \centering
    \includegraphics[width=15cm]{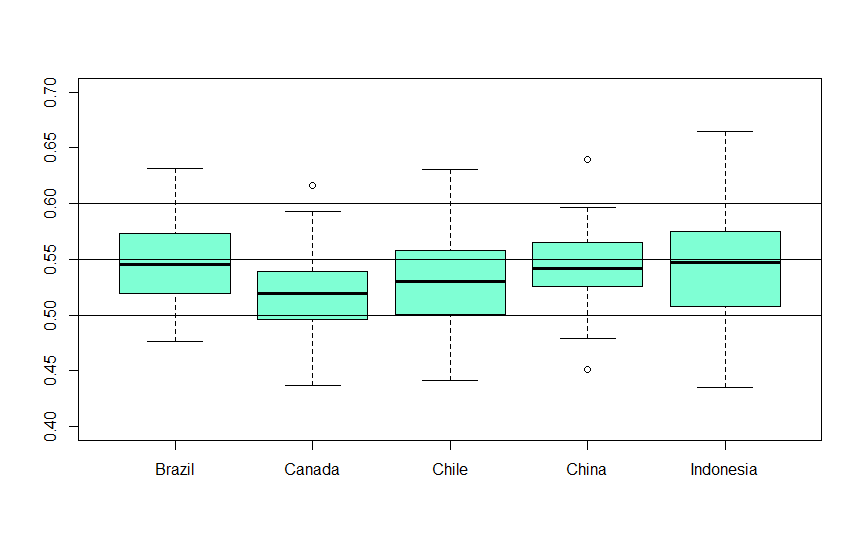}
    \caption{Observed values of Hurst exponent for daily returns in markets of Brazil,Canada,Chile,China and Indonesia }
    \label{fig:Longmem_1}
\end{figure}
\begin{figure}
    \centering
    \includegraphics[width=15cm]{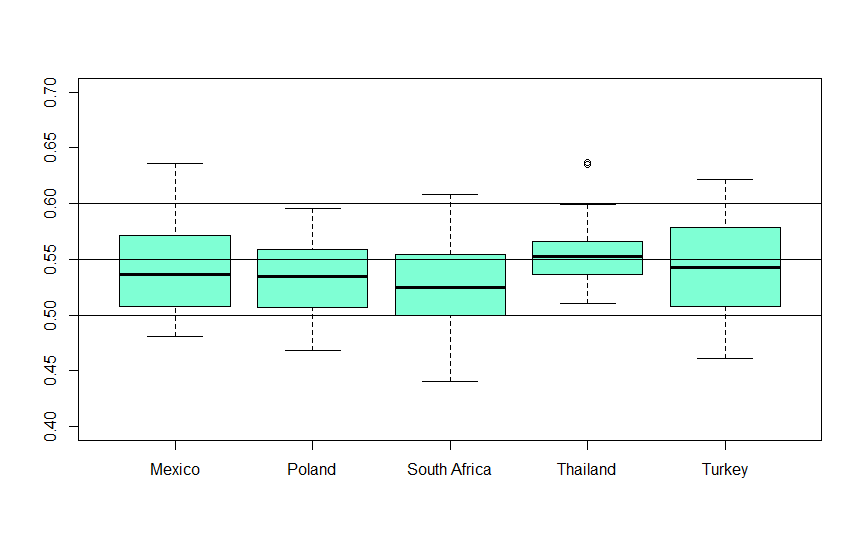}
    \caption{Observed values of Hurst exponent for daily returns in markets of Mexico,Poland,South Africa,Thailand and Turkey}
    \label{fig:Longmem_2}
\end{figure}
Figure \ref{fig:Longmem_1} and \ref{fig:Longmem_2} represents the boxplots of observed hurst exponents of daily return series.It is clearly visible from the boxplots that less than or equal to 50 \% of the data lies in the range 0.55 to 0.6 in every market. Hence,the stylized fact has been contradicted. However,about 70\% of the data lies above 0.5 in every market which establishes the presenece of long memory in daily return series.By visual inspection, it can be said that tendency of exhibiting long memory in daily returns in stocks listed on Toronto Stock Exchange (Canada) is least exchange and tendency of exhibiting long returns in stocks listed on The Stock Exchange of Thailand is maximum among all markets considered.

\subsubsection{Long memory in Volume Series}
Observed hurst exponents of volume series of all the stocks considered were greater than 0.5 as expected. Hence, Volume series does show long memory.

\subsubsection{Slow Decay of autocorrelations in absolute Returns }
\begin{figure}
    \centering
    \includegraphics[width=15cm]{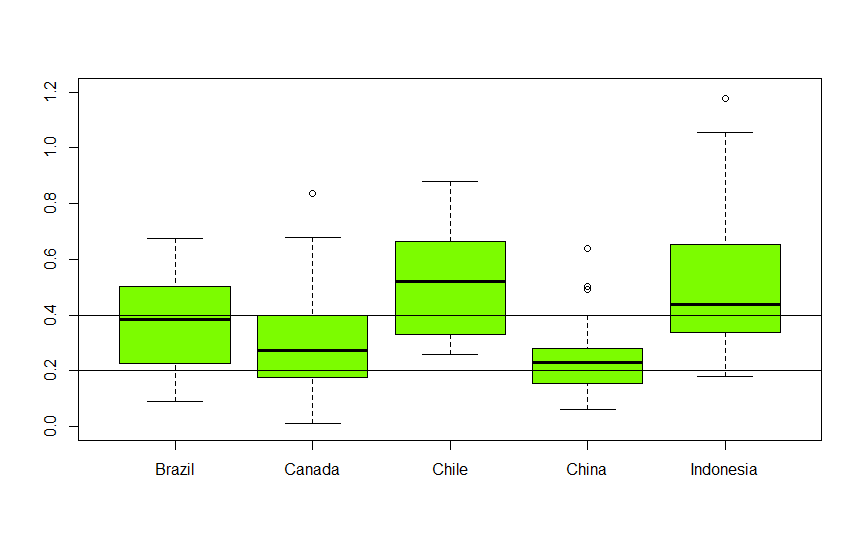}
    \caption{Observed values of Power law exponent for autocorrelations of absolute daily returns in markets of Brazil,Canada,Chile,China and Indonesia}
    \label{fig:exp_1}
\end{figure}
\begin{figure}
    \centering
    \includegraphics[width=15cm]{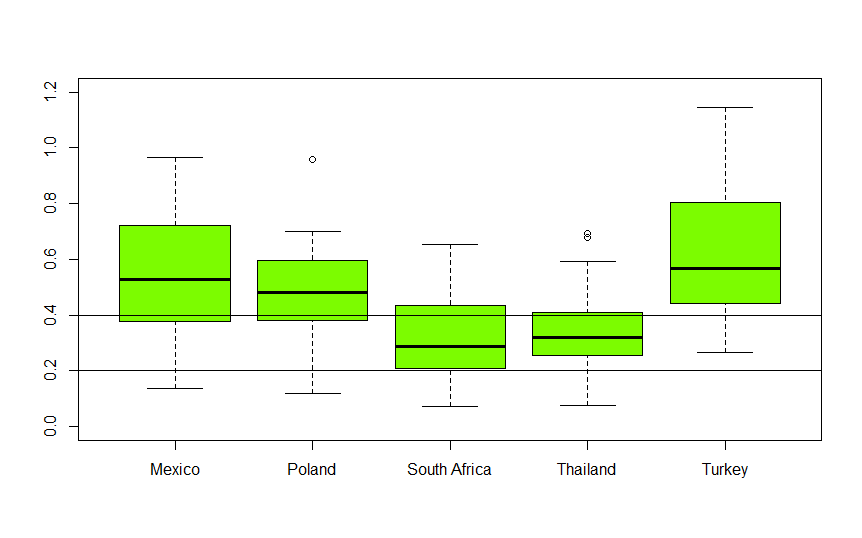}
    \caption{Observed values of Power law exponent for autocorrelations of absolute daily returns in markets of Mexico,Poland,South Africa,Thailand and Turkey}
    \label{fig:exp_2}
\end{figure}
Figure \ref{fig:exp_1} and \ref{fig:exp_2} shows boxplots of observed power law exponents for autocorrelations of absolute daily returns in the range 0 to 1.2. It can be observed that more than 50\% of the data lies outside the expected range [0.2,0.4] in all the markets except Thailand.\par
However, most of the observed values of power law exponents is smaller than 0.4 in case of Canada,China,South Africa and Thailand suggesting a slow decay of autocorrelations.By looking at the boxplot in case of China, it can be said that the decay of autocorrelation is much slower compared to other markets. Most of the observed values of power law exponents are greater than 0.4 in case of Chile, Indonesia, Mexico, Poland and Turkey suggesting fast decay of autocorrelation of absolute returns than expected.The decay of autocorrelations is faster in case of Indonesia and Turkey than the decay of autocorrelations in other markets considered.

\subsubsection{Absence of autocorrelations}
\begin{table}
\centering
\begin{tabular}{|c|c|c|}
\hline
\multirow{2}{*}{\hspace{2em}Country\hspace{2em}} & \multicolumn{2}{c|}{Proportion of stocks for which null hypothesis does not get rejected} \\
\cline{2-3}
 & \hspace{5em}Ljung-Box\hspace{5em} & Box-Pierce  \\
\hline
 Brazil & 0.54 & 0.54\\
 Canada & 0.53 & 0.53\\
 Chile & 0.08 & 0.08\\
 China & 0.11 & 0.11\\
 Indonesia & 0.33 & 0.33\\
 Mexico & 0.31 & 0.31\\
 Poland & 0.30 & 0.30\\
 South Africa & 0.39 & 0.39\\
 Thailand & 0.37 & 0.37\\
 Turkey & 0.57 & 0.57\\
\hline
\end{tabular}
\label{table:5}
\caption{Proportion of stocks for which null hypothesis does not get rejected in Ljung-Box and Box-Pierce test }
\end{table}

Table 5 shows the proportion of stocks for which null hypothesis does not get rejected in Ljung-Box and Box-Pierce test for lag 10 at level of significance 0.05 . It is clearly seen that the proportions are less than 0.5 in all the markets except the markets of Brazil, Canada and Turkey.Even in case of Brazil, Canada and Turkey the proportions are close to 0.5. Hence, it can be said that this stylized fact has been contradicted except in case of Brazil, Canada and Turkey. Very low proportions in case of Chile and China suggests strong presence of autocorrelations in returns. This strong autocorrelation need to be used in prediction of returns in these markets.

\subsubsection{Volatility Clustering}
\begin{table}
\centering
\begin{tabular}{|c|c|c|}
\hline
\multirow{2}{*}{\hspace{2em}Country\hspace{2em}} & \multicolumn{2}{c|}{Proportion of stocks for which null hypothesis gets rejected} \\
\cline{2-3}
 & \hspace{5em}Ljung-Box\hspace{5em} & Box-Pierce  \\
\hline
Brazil & 0.98 & 0.98\\
Canada & 0.85 & 0.85\\
Chile & 0.96 & 0.96\\
China & 1 & 1\\
Indonesia & 0.98 & 0.98\\
Mexico & 0.92 & 0.92\\
Poland & 1 & 1\\
South Africa & 1 & 1\\
Thailand & 1 & 1\\
Turkey & 0.95 & 0.95\\
\hline

\end{tabular}
\label{table:6}
\caption{Proportion of stocks for which null hypothesis gets rejected in Ljung-Box and Box-Pierce Test}
\end{table}

\begin{table}
\centering
\begin{tabular}{|c|c|c|c|c|c|}
\hline
\multirow{2}{*}{Country} & \multicolumn{5}{c|}{Proportion of stocks for which null hypothesis gets rejected} \\
\cline{2-6}
 & \hspace{1em}Lag 1\hspace{1em} & \hspace{1em}Lag 2\hspace{1em} & \hspace{1em}Lag 3\hspace{1em} & \hspace{1em}Lag 4\hspace{1em} & \hspace{1em}Lag 5\hspace{1em}  \\
\hline
Brazil & 0.96 & 0.86 & 0.66 & 0.68 & 0.52\\
Canada & 0.87 & 0.65 & 0.65 & 0.62 & 0.53\\
Chile & 0.96 & 0.92 & 0.88 & 0.67 & 0.58\\
China & 0.97 & 0.95 & 0.97 & 0.97 & 0.97\\
Indonesia & 0.96 & 0.87 & 0.88 & 0.67 & 0.71\\
Mexico & 0.92 & 0.73 & 0.77 & 0.58 & 0.58\\
Poland & 0.96 & 0.74 & 0.52 & 0.52 & 0.65\\
South Africa & 0.94 & 0.90 & 0.77 & 0.68 & 0.61\\
Thailand & 1 & 0.97 & 0.91 & 0.86 & 0.83\\
Turkey & 1 & 0.86 & 0.91 & 0.80 & 0.52\\
\hline
\end{tabular}
\label{table:7}
\caption{Proportion of stocks for which null hypothesis gets rejected when tested for autocorrelation coefficient with certain lag 0 }
\end{table}

Table 6 shows the proportion of stocks for which null hypothesis gets rejected in portmanteau tests with lag 5 at 0.05 level of significance in each market.The high values of proportion suggests the volatility clustering in the data.
\newline
Table 7 shows the proportion of stocks for which null hypothesis gets rejected when tested for autocorrelation coefficient with certain lag 0(Two test statistic can be used to perform this test. However,the proportion is same in all the cases irrespective of the test statistic used.) .All the proportions are greater than 0.5,suggesting non-zero autocorrelation coefficients upto lag 5 in all the markets. Very high value of proprtion in chinese market suggests strong presence of volatility clustering in the market. This property need to be used while modelling the daily return series. Overall, the data validates the presence of volatility clustering in financial time series data.

\subsubsection{Conditional Heavy Tails}
Observed values of tail index for residuals obtained after fitting GARCH(1,1) model to the data of daily returns of stocks are finite except for the five stocks in Chinese market for which the distribution of daily returns did not have heavy tails.
\begin{table}
\centering
\begin{tabular}{ |p{3cm}|p{7.5cm}|  }
\hline
Country & Proportion of stocks for which tail index has decreased after fitting a GARCH model \\
\hline
Brazil & 0.76 \\
Canada & 0.87 \\
Chile & 0.46 \\
China & 0.58 \\
Indonesia & 0.42 \\
Mexico & 0.62 \\
Poland & 0.57 \\
South Africa & 0.81 \\
Thailand & 0.51 \\
Turkey & 0.73 \\

\hline
\end{tabular}
\label{table:8}
\caption{Proportion of stocks for which tail index of residuals obtained is less than the tail index of returns }
\end{table}
\newline
As mentioned in article \cite{Indian_market},the tail index does not decrease in most of the stocks in Indian Market  when a GARCH model is fitted. Table 8 shows the proportion of stocks for which tail index gets reduced after fitting a GARCH model. The proportion is less in case of Chile,Indonesia and Thailand similar to Indian Market. High values of proportion in markets of Canada,South Africa and turkey confirms the reduction of tail index after fitting the GARCH model in these markets. 

\subsubsection{Intermittancy}

\begin{table}
\centering
\begin{tabular}{|c|c|c|}
\hline
\multirow{2}{*}{\hspace{2em}Country\hspace{2em}} & \multicolumn{2}{c|}{Proportion of stocks for which null hypothesis gets rejected} \\
\cline{2-3}
 & \hspace{5em}For returns\hspace{5em} & For residuals obtained after fitteing GARCH model  \\
\hline
Brazil & 0.96 & 0.96\\
Canada & 1 & 1\\
Chile & 1 & 1\\
China & 1 & 1\\
Indonesia & 1 & 1\\
Mexico & 0.85 & 0.85\\
Poland & 0.83 & 0.83\\
South Africa & 1 & 1\\
Thailand & 1 & 1\\
Turkey & 1 & 1\\
\hline
\end{tabular}
\label{table:9}
\caption{Proportion of stocks for which null hypothesis gets rejected when tested if excess kurtosis is 0}
\end{table}

Table 9 shows the proportion of stocks for which null hypothesis gets rejected when one-sided test to decide if excess kurtosis is 0 is performed at 0.05 level of significance. High values of proportion across all the markets for both the data series(Returns and  Residuals obtained after fitting GARCH model to returns series) validates the stylized fact 'Intemittancy' in the returns series. Comparitively, less values of proportion in Polish and Mexican market suggests less intermittent returns in these two markets compared to other markets.

\subsubsection{Taylor effect}
\begin{figure}
    \centering
    \includegraphics[width=15cm]{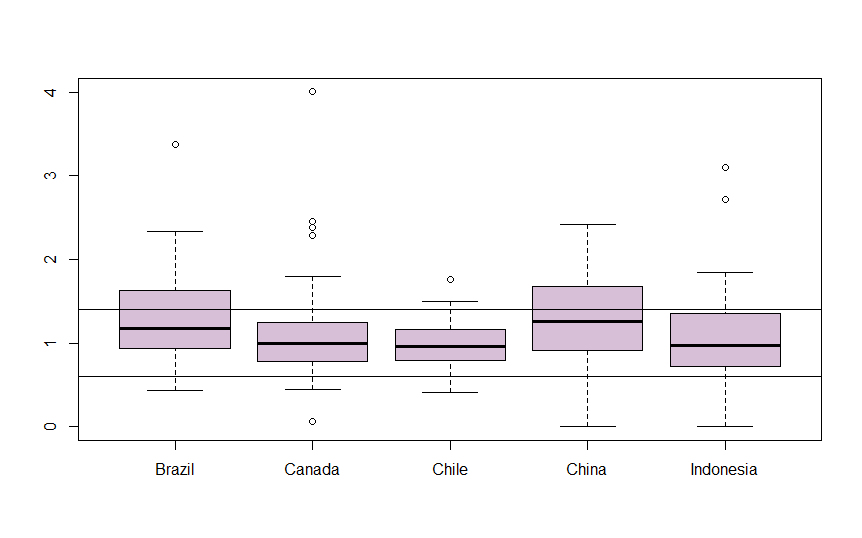}
    \caption{Observed values of d for which lag 1 Autocorrelation of $|Returns|^d$ is maximized in markets of Brazil,Canada,Chile,China and Indonesia}
    \label{fig:tayl_1}
\end{figure}
\begin{figure}
    \centering
    \includegraphics[width=15cm]{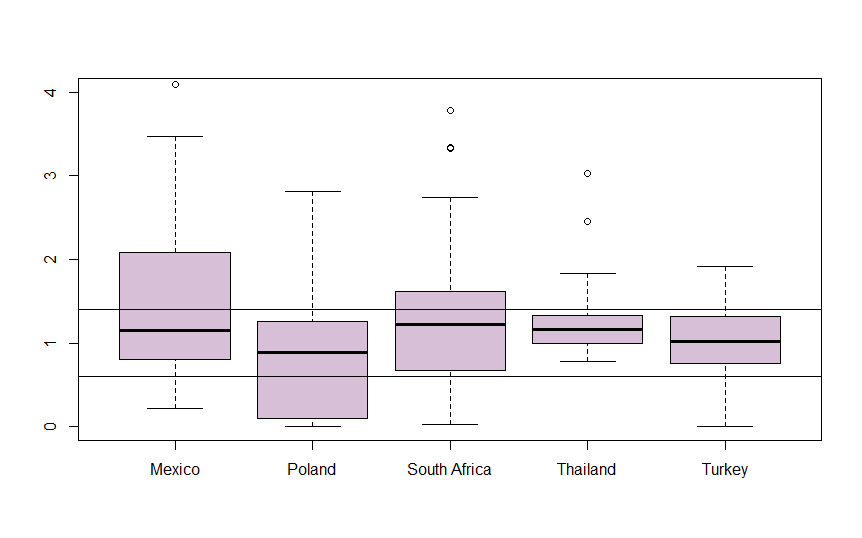}
    \caption{Observed values of d for which lag 1 Autocorrelation of $|Returns|^d$ is maximized in markets of Mexico,Poland,South Africa,Thailand and Turkey}
    \label{fig:tayl_2}
\end{figure}
Figure \ref{fig:tayl_1} and \ref{fig:tayl_2} shows the boxplots of the values of d for which aurocorrelation with lag 1 of $|Returns|^d$ is maximized in the range 0 to 4. Median of observed d values lie close to 1 in all the markets. Majority of the data lies in the interval [0.6,1.4] in markets of Canada, Chile, Indonesia, Thailand and Turkey.Observed values of d are lower in case of Poland and higher in case of Mexico.
\section{Conclusions}
\begin{table}
\centering
\begin{tabular}{ |p{4.5cm}|p{7cm}|p{7cm}|  }
\hline
Stylized Empirical Fact & Verified in & Contradicted in  \\
\hline
Gain Loss Asymmetry & Canada, Poland, South Africa & Chile, Indonesia and Thailand\\
Leverage Effect & Canada, Mexico, Poland and South Africa and   & Chile, China, Indonesia and Thailand\\
Aggregational Gaussinity & Chile, Indonesia, Thailand and Turkey  & None\\
Heavy Tails & All markets & None\\
Decay of Distribution of Volume as Power Law  & All markets & None \\
Volume Volatility Correlation & All Markets  & None\\
Risk return tradeoff & China, Indonesia, Poland, Thailand and Turkey  & Brazil, Canada, Chile, Mexico, South Africa\\
Asymmetry in time scales & All markets & None \\
Long Memory & None  &  All markets\\
Long Memory in Volume Series & All markets  & None\\
Slow Decay of Autocorrelations in absolute returns & Thailand  & None\\
Absence of autocorrelations & Brazil, Canada and Turkey  & Chile, China, Indonesia, Mexico, Poland, South Africa and Thailand\\
Volatility Clustering & All markets  & None\\
Conditional Heavy Tails & Canada, South Africa and Turkey  & Chile, Indonesia and Thailand\\
Intermittancy & All markets  & None\\
Taylor effect & All markets  & None\\
\hline
\end{tabular}
\label{table:10}
\caption{Summary of Data analysis }
\end{table}
Table 10 presents the summary of data analysis. 17 stylized empirical facts have been considered for data analysis. A 17 dimensional vector can be associated with each market, where $i^{th}$ component of the vector is 1,-1 if $i^{th}$ stylized fact has been verified or contradicted respectively in the corresponding market. If the stylized fact is neither contradicted nor verified in the corresponding market, the corresponding component of the vector assumes value 0. Note that, -1,0,1 are considered to be ordinal categorical variables.
\begin{figure}
    \centering
    \includegraphics[width=15cm]{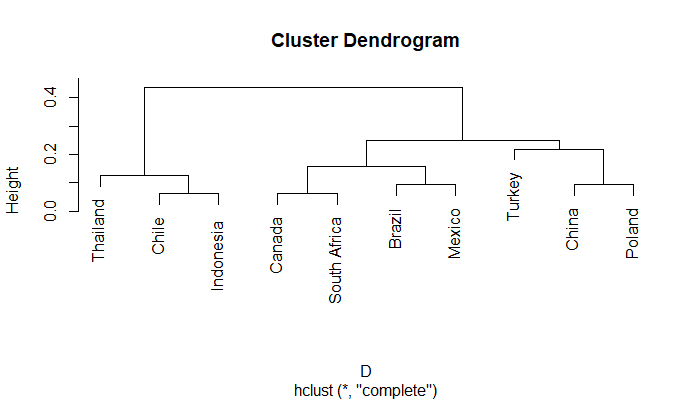}
    \caption{Dendrogram}
    \label{fig:dendrogram}
\end{figure}
The hierarchical clustering method was applied on the data. Figure \ref{fig:dendrogram} is the dendrogram obtained after clustering was carried out. It can be seen that Chile, Indonesia and Thailand forms the one cluster. Comparing with results corresponding to Indian market presented in article \cite{Indian_market}, it can be said that behaviour of Indian market is similar to Thai and Indonesian market. Other two clusters formed are Canada, South Africa, Brazil, Mexico and Turkey, China, Poland.

\section*{Acknowledgements}
The work presented in this paper was completed when one of the authors Mr. Vaibhav Sherkar was working under guidance of Dr. Rituparna Sen as a Summer research fellow selected through summer research fellowship program by Indian Academies of Sciences (IASc) Bengaluru, Indian National Science Academy (INSA) New Delhi and National Academy of Sciences,India (NASI) Prayagraj. We are thankful to them. We are grateful to Indian Statistical Institute, Bengaluru for providng necessary infrastructural support. 
\newpage
\printbibliography

@article{cont,
  author =       "Rama Cont",
  title =        "Empirical properties of asset returns:stylized facts and                          statistical issues",
  journal = "Quantitative Finance" ,
  volume = "1",
  pages = "223-236",
  year = "2001"
}

@article{longmemory,
    author = "G.Pernagallo and B.Torrisi",
    title = "An empirical analysis on the degree of gaussinity and long memory of financial returns in emerging economies",
    journal = "PhysicaA(2019)",
    volume = "527",
    DOI = "https://doi.org/10.1016/j.physa.2019.121296"
}

@article{Asymmetry,
    author = "Ulrich A.Müller and Michel M.Dacorogna and Rakhal D.Davé and Richard B. Olsen and Olivier V. Pictet and Jacob E. von Weizsäcker",
    title = "Volatilities of different time resolutions -Analyzing the dynamics of market components",
    journal = "Journal of Empirical Finance",
    volume = "4",
    pages = "213-239",
    year = "1997"
}

@article{Taylor,
    author = "Scott Thompson",
    title = "The Stylised Facts of Stock Price Movements",
    journal = "New Zealand Review of Economics and Finance",
    volume = "1",
    pages = "50-77",
    year = "2013"
}

@article{ Indian_market,
    author = " Rituparna Sen, Manavathi Subramaniam " ,
    title = "Stylized Facts of the Indian Stock Market",
    journal = "Asia Pacific Financial Markets",
    volume = "26",
    pages = "479-493",
    DOI = "https://doi.org/10.1007/s10690-019-09275-3",
    year = "2019"
}

@book{Time_series,
  author = {Ruey S. Tsay},
  title = {Analysis of financial time series},
  publisher = {Wiley},
  edition = {2}
}

@article{UCL_article,
    author = "Martin Sewell",
    title = "Characterization of Financial Time Series",
    year = "2011"
}

@article{ Morocco,
    author = "  Moulay Driss Elbousty,  Lahsen Oubdi " ,
    title = " Volatility stylized facts in the Moroccan stock market: Evidence from both aggregate and disaggregate data",
    journal = " Turkish Economic Review",
    volume = "7",
    issue = "2",
    DOI = " http://dx.doi.org/10.1453/ter.v7i2.2077",
    year = "2020"
}

@article{ Nigeria,
    author = "  Wilson E. Herbert,  Georgina O. Ugwuanyi,Ernest I. Nwaocha " ,
    title = " Volatility Clustering, Leverage Effects and Risk-Return Trade-Off in the Nigerian Stock Market",
    journal = "Journal of Finance and Economics ",
    volume = "7",
    issue = "1",
    pages = "1-13",
    DOI = " 10.12691/jfe-7-1-1",
    year = "2019"
}

@article{ Peru,
    author = " Alberto Humala ,Gabriel Rodríguez   " ,
    title = " Some Stylized Facts of Returns in the Foreign Exchange and Stock Markets in Peru",
    journal = "Studies in Economics and Finance ",
    volume = "30",
    issue = "2",
    DOI = " 10.1108/10867371311325444",
    year = "2013"
}
\end{document}